\documentclass[fleqn]{elsart}
\usepackage{cite}
\usepackage{subfigure}
\usepackage{rotate}
\usepackage{epsfig}
\usepackage{amsmath}

\newcommand{\mean}[1]{\left\langle #1 \right\rangle}
\newcommand{\bm}[1]{\vec{#1}}

\journal{Physica A}
\begin{document}

\begin{frontmatter}
  \title{Advantages of Hopping on a Zig-zag Course}

  \author[hub]{Lutz Schimansky-Geier},
  \ead{alsg@physik.hu-berlin.de}
  \author[hub]{Udo Erdmann\corauthref{cor1}}
  \ead{udo.erdmann@physik.hu-berlin.de}
  \corauth[cor1]{Corresponding author.}
  \author[hub]{Niko Komin}
  \address[hub]{Institute of Physics, Humboldt-University Berlin, 12489
    Berlin, Newtonstr. 15, Germany}

  \begin{abstract}
    We investigate self-moving particles which prefer to hop with a certain
    turning angle equally distributed to the right or left. We assume this
    turning angle distribution to be given by a double Gaussian distribution.
    Based on the model of Active Brownian particles and we calculate the
    diffusion coefficient in dependence on the mean and the dispersion of the
    turning angles. It is shown that bounded distribution of food in patches
    will be optimally consumed by the objects if they hop preferably with a
    given angle and not straight forwardly.
  \end{abstract}
\end{frontmatter}

\section{Introduction}
In 1827, when the British botanist {Robert Brown} discovered the erratic motion
of small particles immersed in a liquid, he considered them first as living
entities. Later on Einstein and Smoluchowski \cite{EiSm97} have shown that the
behavior of Brownian particles are due to the molecular agitation, i.e.
collisions from the side of the solvent molecules only.

The most fascinating formulation was given by Paul Langevin \cite{La08} who
added random forces to the equation of motion, in addition to the viscous
forces acting on the particles in the fluid. Recently, this latter formulation
was used to study active particles which are self-moving particles thanks to
some energy supply and, thus, remember the entities as they were considered by
Brown. This concept of active Brownian particles is easy, negative friction at
smaller velocities resembles a pump of energy. This energy is transformed into
kinetic one up to some velocity (for higher ones, dissipation wins again) and
the particles starts to be moving along their given direction. In case of
additionally acting forces, in particular random ones, the individuals change
the direction as well as the modulus of their velocity.

Several interesting phenomena were studied in the past. Simple models of
active particles were studied already in many earlier works
\cite{SchiGr93,HeMo95,EbSchwTi99,ErEbSchiSchw99}. For broader reviews one
might refer to \cite{He01,MiCa02,ErEbSchiOrMo03}. Other previous versions of
Active Brownian particle models \cite{SchiMiRoMa95,Er99} consider more
specific activities, such as environmental changes and signal--response
behavior. In these models, the Active Brownian particles (or active walkers,
within a discrete approximation) are able to generate a self-consistent field,
which in turn influences their further movement and physical or chemical
behavior.  This non-linear feedback between the particles and the field
generated by themselves results in an interactive structure formation process
on the macroscopic level. Hence, these models have been used to simulate a
broad variety of pattern formations in complex systems, ranging from physical
to biological and social systems \cite{SchiMiRoMa95,HeSchwKeMo97,ShChPa99}.

The concept of Active Brownian particles was also applied to the motion of
small water flees, so called {\em Daphnia} \cite{ErEbSchiOrMo03,MaSchw03}
where it can also explain some collective swarming and curling of the animals
and in light shafts \cite{OrBaMo02b,ErEb02}. These {\em Daphnia} exhibit
another property during their motion as was recently detected in measurements
\cite{OrBaMo02b}.  During a small time step they hop with some preferred angle
to the left or right of their previous motion.  Hence the direction of motion
is due to some given distribution of turning angles which determines the
preferred direction after the time step.

In \cite{KoErSchi03} we applied a random walk model to find the effective
diffusion coefficient of the random motion with preferred turning angles. For
this purpose we have used and expression derived by Kareiva and Shigesada (see
\cite{OkLe02engl} for details) and applied it to the experimental observations
\cite{OrBaMo02b}. Here we develop a continuous model which is based on the
stochastic equations of motion in two dimensions. Again we find a decrease of
diffusion with the mean turning angle and a non-monotonous dependence on the
dispersion of angles.

In addition we let the particle consume some non-moving food with constant
rates. If this food is distributed in bounded regions, it would be suggestive
to hop with a preferred turning angle. That's why the food consumption in a
given time is maximized as will be shown in the last section.

\section{Equations of motions of Active Brownian particles}

The heart of Active Brownian particles is an energy pump modeled by negative
dissipation at small velocities. Thus we suppose that the relaxation
coefficient of the velocities is negative for small velocities. For large
velocities dissipation again overwhelms. A good prototypical choice goes back
to Rayleigh who considered first nonlinear friction as \cite{Ra94engl}
\begin{equation}
  \label{eq:friction}
  \gamma(v)\,=\,- \alpha \,+\, v^2
\end{equation}
where $v$ is the modulus of the velocity-vector. This type of friction
function describes a source of energy if $v^2 < \alpha$ and lowers the
velocity in the opposite case. This is done upto a stationary velocity which
has the absolute value of $\sqrt{\alpha}$.

Since {\em Daphnia} hop in first approximation in a two dimensional plane we
assume that the position and the velocities are determined by the equations of
motion:
\begin{equation}
  \label{langev-or}
  \frac{{\rm d} \bm{r}}{{\rm d} t}=\bm{v}\;,\qquad
  \frac{{\rm d} \bm{v}}{{\rm d} t}= -\gamma(v) \bm{v}
  +\bm{\omega} \times \bm{v} + \sqrt{2 D_v}\,{\bm{\xi}}(t)
\end{equation}
We assume that $\bm{r}=\{r_1,r_2\}$ and $\bm{v}=\{v_1,v_2\}$ are
two-dimensional vectors, respectively, for the position and the velocity of
the considered particle. $D_v$ scales the randomness of the motion which is
due to the stochastic sources in (\ref{langev-or}). Therein ${\bm{\xi}}(t)$ is
a Gaussian stochastic force with unit strength, independent components and a
$\delta$-correlated time dependence
\begin{equation}
  \label{stoch}
  \mean{\bm{\xi}(t)}=0 \;,\qquad
  \mean{{\xi}_i(t){\xi}_j(t')}= \, \delta_{ij} \, \delta(t-t').
\end{equation}

The new term in (\ref{langev-or}) is the Lorentz-like force which models the
circular motion. We assume that $\bm{\omega}$ is directed perpendicularly to
the plane of motion, i.e.  $ \bm{\omega}=\{0,0,\omega\}$ and a Larmor-like
rotation is generated.  The sign of $\omega$ determines clockwise or counter
clockwise motion.  Later we will assume that half of a population is running
clockwise and the second half counter clockwise.

The situation can be better understood if we change to polar
coordinates in the velocity space by introducing
\begin{equation}
  \label{eq:polar_veloc}
  v_1=v(t)\,\cos\left(\phi(t)\right)\;,
  \qquad v_2=v(t) \,\sin\left(\phi(t)\right)
\end{equation}
This change of variables is quite similar to an often used
transformation in the phase space of nonlinear oscillators
\cite{HaRi83}. In contrast we consider a transformation within the
two-dimensional velocity space.

The equations of motion transform into
\begin{subequations}
\label{eq:polar}
\begin{eqnarray}
  \label{eq:mod_velo}
  {{\rm d}v\over {\rm d}t}  &=& -\gamma(v)\,v + \cos(\phi)\, {\xi}_1
  + \sin(\phi) \,{\xi}_2\\
  \label{eq:angle_velo}
  {{\rm d} \phi \over {\rm d}t} &=& \omega + {1 \over v}
  \left(-\sin(\phi) \,{\xi}_1  +  \cos(\phi)\, {\xi}_2\right).
\end{eqnarray}
\end{subequations}
Without noise the velocity vector relaxes to a constant modulus
$v=\sqrt{\alpha}$ which moves with constant angular frequency $\omega$.

In Stratonovich-interpretation the corresponding Fokker-Planck equation for
the density in the $v,\phi$-space then follows \cite{St67,AnAsNeVaSchi02,FPE}
\begin{equation}
  \label{eq:fpe_velo}
    \frac{\partial}{\partial t} P(v,\phi,t)
    = \frac{\partial}{\partial {v}}\left[\gamma({v})\,{v}
      -{D_v \over v}\right]\, P  -\omega\,\frac{\partial P}{\partial {\phi}}
    + D_v\, \frac{\partial^2 P}{\partial {v^2}}
    + {D_v\over v^2}\, \frac{\partial^2 P}{\partial {\phi^2}}.
\end{equation}
Its stationary distribution can be easily found
\begin{equation}
  \label{eq:distr_gen}
  P_0\,(v, \phi)= {N} \,v \,\exp\left(-{1\over D_v}\,
    \int \gamma(v)\,v\,{\rm d}v\right),
\end{equation}
which is uniformly distributed in $\phi$. We see that the rotational motion
has no influence on the stationary density.  We mention that the Larmor
rotations are inducing a circulating probability flow. The most probable
values can be easily found from (\ref{eq:distr_gen})
\begin{equation}
  \label{eq:most_prob}
  \gamma(\tilde{v})-{D_v \over \tilde{v\,}^2 }=0\,.
\end{equation}

In the limit of weak noise and strong pumping $D_v/\alpha \rightarrow 0$ we
approximate the velocity distribution by a $\delta$-function around the most
probable values ($v \ge 0,\tilde{v}>0$)
\begin{equation}
  P_0\,({v}, \phi) = {1\over 2 \pi}\delta\left( v - \tilde{v}\right)\,.
\end{equation}

\section{Mean square displacement}

We will characterize the motion of the active particles by its mean square
displacement. In our case without external confinements the random motion
induces linear diffusion in the long time limit at large length scales. Hence
the mean square displacement is determined by the diffusion coefficient which
is in two dimensions
\begin{equation}
  \left\langle \left(\bm{r}(t) -\bm{ r}(0)\right)^2 \right\rangle
  = 4\, D_r \,t\,.
\end{equation}
For a more detailed discussion of linear diffusion and the
connected Einstein relations see \cite{HaTh82,HTB90}.

In order to calculate $D_r$ we integrate the velocity, which yields
\begin{equation}
  \bm{r}(t) = \bm{r}(0) + \int_0^t {\rm d} t_1 \bm{v}(t_1).
\end{equation}
Therein we agreed that the particles have been started at $t = 0$ in
$\bm{r}(0)$. By this way the mean square displacement reads
\begin{equation}
  \left\langle \left(\bm{r}(t) -\bm{r}(0)\right)^2 \right\rangle
  = \int_0^t {\rm d} t_1 \int_0^t {\rm d} t_2
  \left\langle \bm{v}(t_1) \bm{v}(t_2) \right\rangle,
\end{equation}
and is traced by to the velocity correlation function. Crossing to radial and
angle variables in velocity space (Eq.(\ref{eq:polar_veloc})) gives
\begin{equation}
  K(t) = \left\langle \bm{v}(t_1) \bm{v}(t_2) \right\rangle
  =\left\langle v(t_1)v(t_2) \cos(\phi(t_1)-\phi(t_2) \right\rangle\,.
\end{equation}
Our main approximation is a widely used decoupling of $v(t)$ from the
dynamics of the angle $\phi(t)$ \cite{HaJu95}. Then for the correlation
function we find
\begin{equation}
  K(t)  \simeq \left\langle v(t_1)v(t_2) \right\rangle
  \left\langle \cos(\phi(t_1)-\phi(t_2)) \right\rangle
\end{equation}
More restrictively we require that the modulus of the velocities is
approximately fixed which results in $\langle v(t_1)v(t_2)\rangle =
\tilde{v}^2$ and $\tilde{v}$ is due to Eq.~(\ref{eq:most_prob}).

Hence the mean square displacement is due to the correlations of angles of the
velocity vector. From Eq.~(\ref{eq:fpe_velo}) we get for the
angle-distribution the Fokker-Planck equation
\begin{equation}
\label{eq:fpe_angle}
  \frac{\partial }{\partial t} P(\phi, t)
  = -\omega\,\frac{\partial P}{\partial {\phi}}
  + D_{\phi} \frac{\partial^2 f}{\partial \phi^2}
\end{equation}
with the diffusion coefficient $D_{\phi} = D_v /\tilde{v}^2$. Solving
(\ref{eq:fpe_angle}) with $t_1> t_2$ gives
\begin{equation}
  P(\phi_1, t_1  | \phi_2, t_2) = \frac{1}{\sqrt{4 D_{\phi}\pi (t_1-t_2)}}
  \exp \left[
    -\frac{(\phi_1-\omega(t_1-t_2)-\phi_2)^2}{4 D_{\phi}(t_1-t_2)}\right]\,.
\end{equation}
This expression is used for averaging the cosine of the increment
in angles $\phi(t_1)-\phi(t_2)$. Afterwards the time integrals can
be calculated by standard calculations and a lengthy expression is
obtained. We give here the long time asymptotic behavior where a
linear increase becomes dominant over decaying and constant items.
It has a simple shape as
\begin{equation}
  \label{eq:square_displacement}
  \left\langle (\bm{r}(t) -\bm{ r}(0))^2 \right\rangle
  = 2\, \tilde{v}^2 \frac{D_{\phi}}{D_{\phi}^2
    +\omega^2}\,t + O\left({\rm const},\cos(\omega t),\exp(-D_{\phi}t) \right)
\end{equation}
Hence the diffusion coefficient becomes
\begin{equation}
  \label{eq:diffusion}
  D_r ={\tilde{v}^4 \over 2 D_v}\,
  \left(\frac{1}{1 + \omega^2 \, \tilde{v}^4/D_v^2}\right).
\end{equation}
The first factor agrees with an expression derived by Mikhailov and Meink\"ohn
\cite{MiMe97} in case there is no preference in the turning angle ($\omega =
0$). The second factor in the brackets decreases in case of $\omega \ne 0$.
This solution for the spatial diffusion coefficient is in qualitative agreement
with previous results based on a random walk model \cite{KoErSchi03}.

We point out that the diffusion coefficient does not depend on the sign of
$\omega$, hence not on the direction of rotations. Therefore the derived
expression is true also if part of the population hops clockwise and the other
part counterclockwise.

\begin{figure}[htbp]
  \begin{center}
    \subfigure[]
    {
      \epsfig{file=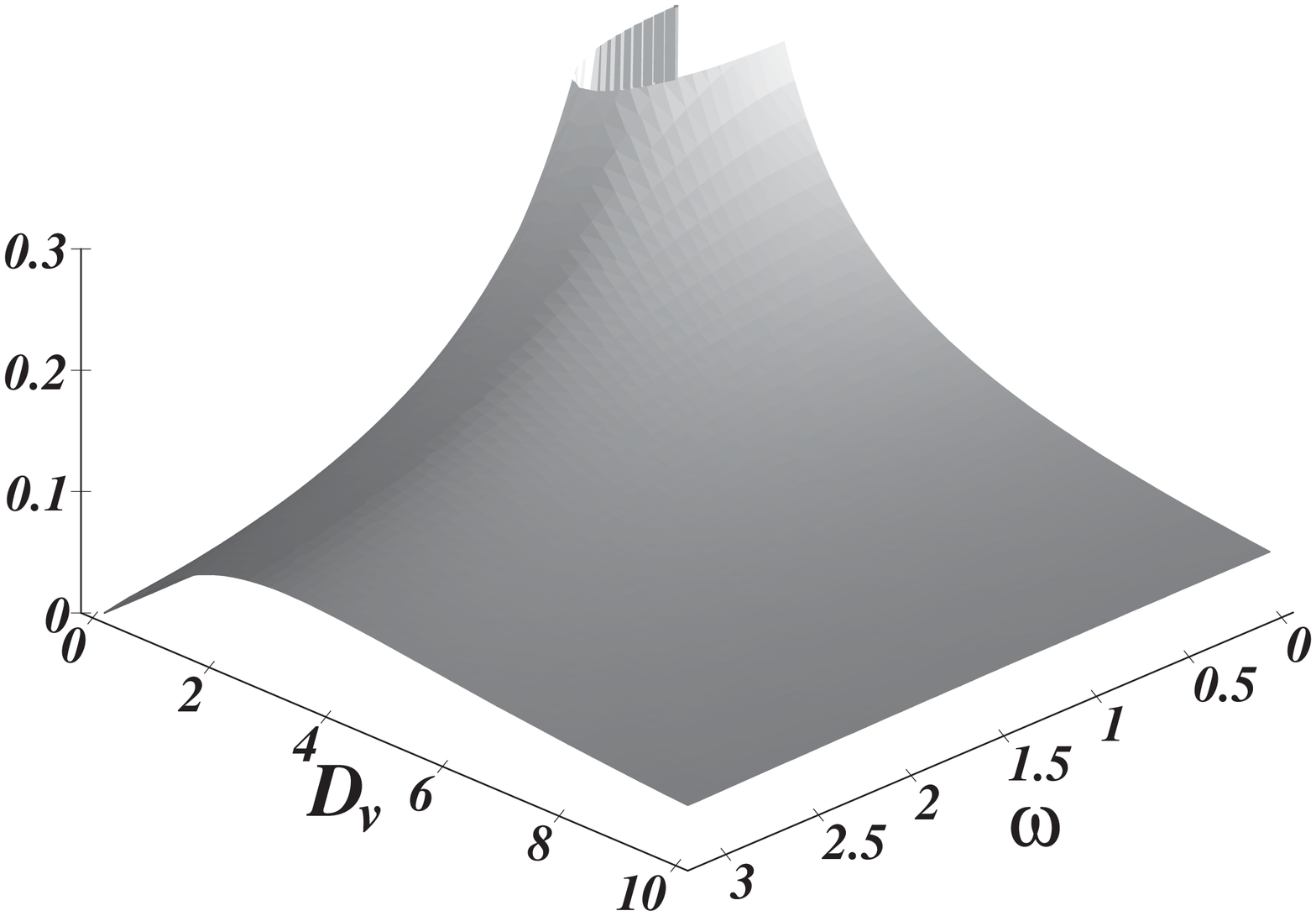,width=6.5cm}
    }
    \subfigure[]
    {
      \epsfig{file=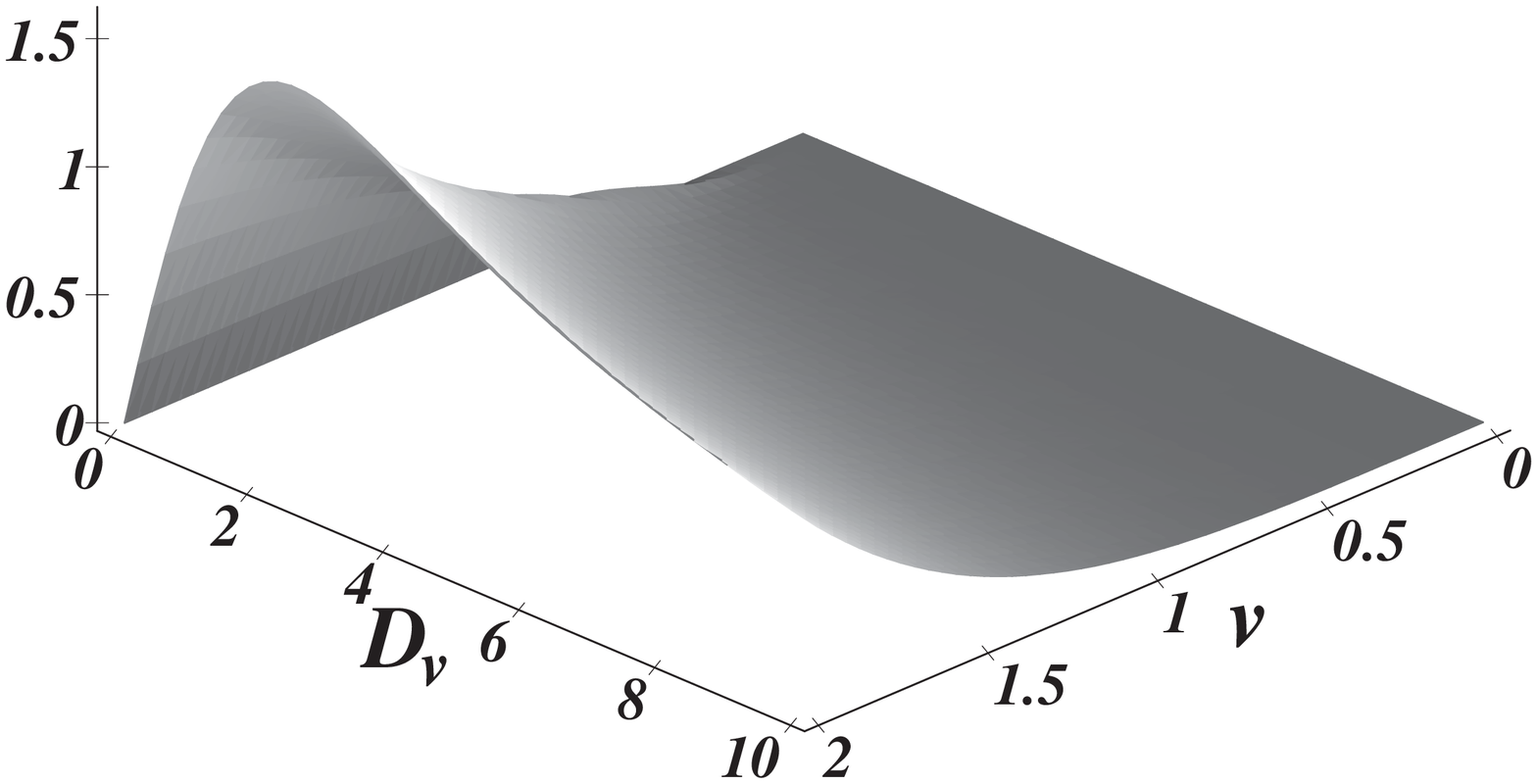,width=6.5cm}
    }
    \caption{Spatial Diffusion $D_r$ depending on external noise $D_v$ and the
      preferred turning angle $\omega$ (a) or the velocity $v$ in the
      stationary case (b)}
    \label{fig:diffusion}
  \end{center}
\end{figure}

The general dependence of the diffusion coefficient~(\ref{eq:diffusion}) on
noise $D_v$ and mean turning angle is shown in Figure~\ref{fig:diffusion}. It
decreases monotonically with the mean turning angle $\omega$. Remarkable one
finds a maximum with respect to the noise. It is located at noise values
\begin{equation}
  \label{eq:noise}
  D_v^{\rm max}\,=\, \tilde{v}^2\,\omega.
\end{equation}
and the spatial diffusion coefficient equals $\tilde{v}^2/4 \omega$.

\section{Optimal food consumption of finite food patches}
\label{sec:optim-food-cons}

In this section we want to develop a simple approach how much food a single
{\em Daphnia} could consume while it is on its foraging path. Assuming that
the density of the animal $\varrho(\bm{r},t)$ obeys a diffusion equation with
the afore-calculated diffusion coefficient $D_r$
\begin{equation}
\label{eq:diff_daphnia}
  \frac{\partial \varrho}{\partial t}
  = D_r \Delta\varrho.
\end{equation}
In two dimensions the resulting probability density in space is
\begin{equation}
  \label{eq:density}
  \varrho(\bm{r},t)=\frac{1}{4\pi\,D_r\,t}
  \exp{\left[-\frac{\bm{r}^{\,2}}{4D_rt}\right]} \, .
\end{equation}
We assume that the particle during its random motion consumes with constant
rate $k$ food $C$ which is given by its density $c(\bm{r},t)$. If the food is
un-movable in space, the consumption of food is described by the dynamics
\begin{equation}
  \label{eq:food}
  \frac{\partial}{\partial t} c(\bm{r},t)= -\,k\,c(\bm{r},t)\,\varrho(\bm{r},t)
\end{equation}

The latter equation can be solved exactly if the solution (\ref{eq:density})
is inserted. Simple quadrature gives
\begin{equation}
  \label{eq:food2}
  c(\bm{r},t) = c_0(\bm{r})\,\exp\left[ -
    \int_{t_0}^t \,k\, \varrho(\bm{r},t^\prime)\, {\rm d}t^\prime\right].
\end{equation}
With the definition of the exponential integral \cite{AbSt84}, $ {\rm E}_1(a)
= \int_a^\infty {e^{-t} \over t }{\rm d}t $, we obtain in compact shape
\begin{equation}
  \label{eq:compact}
  c(\bm{r},t)\,=\,c_0(\bm{r})\,\exp\left[ - {k \over 4 \pi D_r} \,
    {\rm E}_1 \left({\bm{r\,}^2 \over 4D_r t} \right)\right].
\end{equation}
where we have put $t_0=0$.

Let us look at bounded distributed food patches now. As a prototypical
distribution we take spheres with radius $R$ where $C$ is present. Outside of
the circle no food is located. For simplicity we assume the circle positioned
at the origin, i.e.
\begin{equation}
  \label{eq:patch}
  c_0(\bm{r})\,=\, c_0 \quad {\rm if} \quad |\bm{r}| < R.
\end{equation}
and vanishing, elsewhere. Then after a fixed time $T$ we determine the overall
food which is still left. It results from integration
\begin{equation}
  \label{eq:food3}
  C(T) = 2\pi\,c_0 \int_0^R\exp\left[ - {k \over 4 \pi D_r} \,
    {\rm E}_1 \left({r^2 \over 4D_r T} \right)\right] \,r\,{\rm d}r.
\end{equation}
Inspection of this expression shows that there exists a minimum with respect
to the diffusion coefficient $D_r$. Since $D_r$ depends monotonously on the
mean turning angle $\omega$ of the zig-zag hopping it implies a minimum with
respect to the mean angle, too. Thus, the variation of the mean angle with
change of the effective diffusion coefficient gives rise to a change of the
food consumption. One might speculate that nature has found by this way the
reason to optimize the zig-zag hopping.

\begin{figure}[htbp]
  \centering
  \epsfig{file=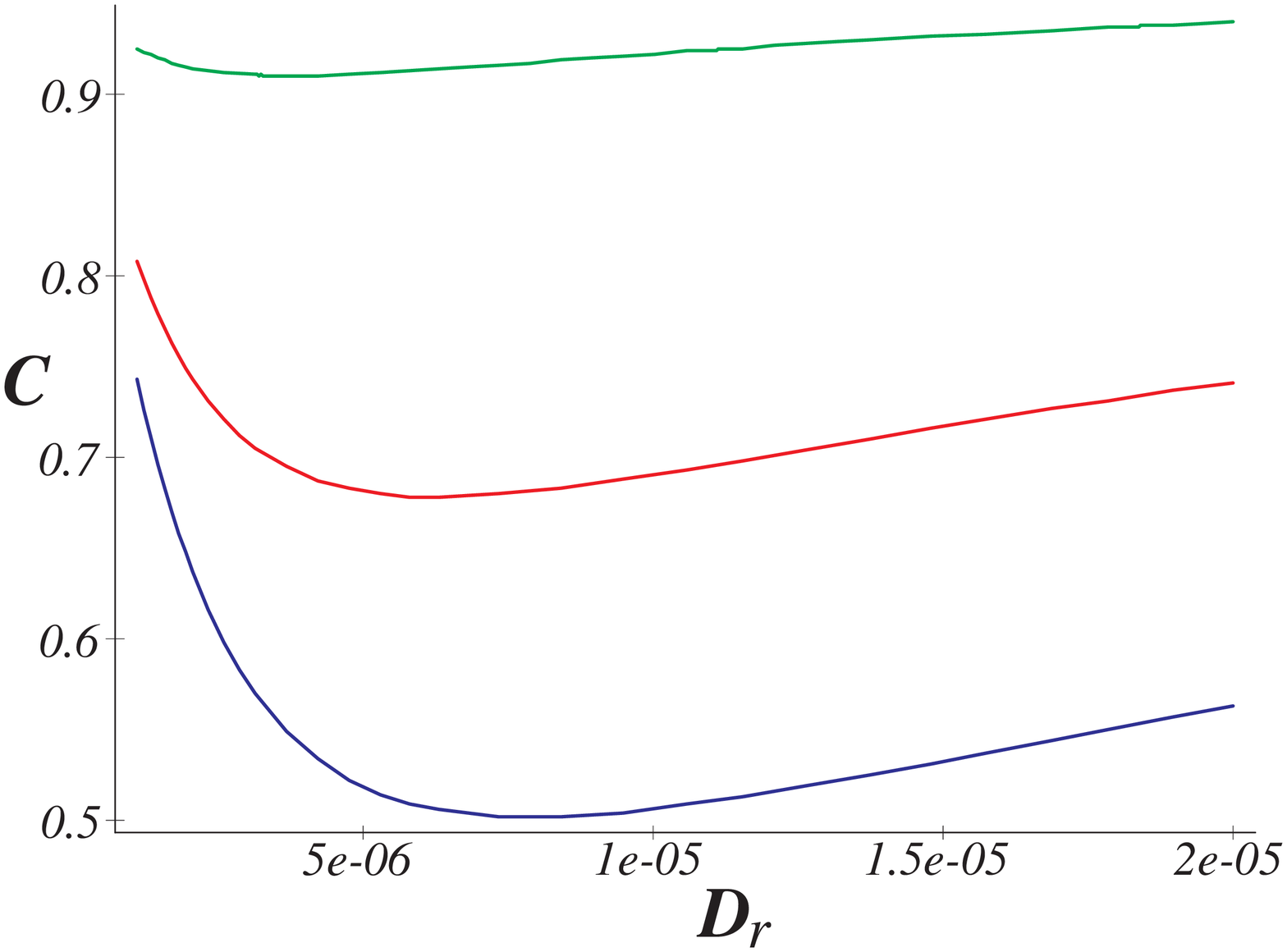,width=10cm}
  \caption{Consumed food depending on the diffusion coefficient.}
  \label{fig:food_consumption}
\end{figure}

The consumed food as a function of the spatial diffusion coefficient $D_r$ is
shown in Fig.\ref{fig:food_consumption}. A simple explanation can be given as
follows. If particles hop straight forwardly they will leave the food region,
going forwardly and backwardly is also of disadvantage. Hence there is a
optimal angle which the particle spends most time in the food patch with. It
does not matter whether they start at the center of the food or not.

\section{Conclusions}

This paper extends the theory of Active Brownian particles with terms which
obey the motion of individuals with preferred turning angle. This motion
could have been observed, for example, in zooplankton populations like {\em
  Daphnia} \cite{OrBaMo02b}. The preferred turning angle distribution, which
could have been easily put into a Random Walk model \cite{KoErSchi03}, comes
into play as a Lorentz-like force within the Langevin equation. This
translation of the preferred turning angle distribution into the concept of
stochastic differential equations makes the treatment of such biological
systems more handy in terms of physical interpretations of the influencing
forces. With the help of the, now more generalized, theory of Active Brownian
particles a broader variety of swarming systems can be described via physical
models.

We calculated the spatial diffusion coefficient, and what is astonishing, is a
maximum of spatial diffusion with the noise strength for preferred turning
angles $-\pi/2<\omega<\pi/2$. The diffusion law (see Eq.~(\ref{eq:diffusion}))
is an extension of the theory presented in \cite{MiMe97}. This result can be
concluded as follows: With a preferred turning angle a individual can cover a
larger area within a given time. This would be an evolutionary advantage as
could have been seen in Sec.~\ref{sec:optim-food-cons}. Here we develop a
theory of food consumption within a bounded food patch. Assuming a Gaussian
distributed density for a single individual, and using the results from the
sections before, a minimum in the left over food with the spatial diffusion
coefficient has been found.  Hence, to cover the maximum of a given area going
zig-zag is many times more beneficial than just normal diffusion. Within a
given time a individual can cover a bigger area with preferred turning angle
and therefore consume more food, which is of evolutionary advantage.

\ack

This work was partly supported by the Collaborative Research Center ``Complex
Nonlinear Processes'' of German Science foundation (DFG-Sfb555). We thank
F.~Moss, I.~M.~Sokolov and W.~Ebeling for fruitful discussions.


\begin{thebibliography}{10}
\expandafter\ifx\csname url\endcsname\relax
  \def\url#1{\texttt{#1}}\fi
\expandafter\ifx\csname
urlprefix\endcsname\relax\def\urlprefix{URL }\fi

\bibitem{EiSm97} A.~Einstein, M.~v. Smoluchowski, Untersuchungen \"uber die
  Theorie der Brownschen Bewegung, 3rd Edition, Vol. 199/207 of Ostwalds
  Klassiker der exakten {N}aturwissenschaften, Harri Deutsch, Frankfurt am
  Main, 1997.

\bibitem{La08} P.~Langevin, C. R. Acad. Sci.  (Paris) 146 (1908) 530--533.

\bibitem{SchiGr93}
M.~Schienbein, H.~Gruler, Bull. Math. Biol. 55 (1993) 585--608.

\bibitem{HeMo95} D.~Helbing, P.~Moln{\'a}r, Phys.  Rev. E 51 (1995)
  4282--4286.

\bibitem{EbSchwTi99} W.~Ebeling, F.~Schweitzer, B.~Tilch, BioSystems 49 (1999)
  17--29.

\bibitem{ErEbSchiSchw99} U.~Erdmann, W.~Ebeling, F.~Schweitzer,
  L.~Schimansky-Geier, Eur. Phys. J. B 15 (2000) 105--113.

\bibitem{He01} D.~Helbing, Rev.  Mod. Phys. 73 (2001) 1067--1141.

\bibitem{MiCa02}
A.~S. Mikhailov, V.~Calenbuhr, From Cells to Societies, Springer,
Berlin, 2002.

\bibitem{ErEbSchiOrMo03} U.~Erdmann, W.~Ebeling, L.~Schimansky-Geier,
  A.~Ordemann, F.~Moss, Active brownian particle and random walk theories of
  the motions of zooplankton: Application to experiments with swarms of {\em
    Daphnia}, \texttt{http://arxiv.org/abs/q-bio.PE/0404018} (Apr.  2004).

\bibitem{SchiMiRoMa95} L.~Schimansky-Geier, M.~Mieth, H.~Ros{\'e}, H.~Malchow,
  Phys. Lett. A 207 (1995) 140--146.

\bibitem{Er99} U.~Erdmann, {\em Interjournal} Complex Systems (1997) 114.
\newline\urlprefix\url{http://www.interjournal.org/manuscript\_abstract.php?10%
536}

\bibitem{HeSchwKeMo97} D.~Helbing, F.~Schweitzer, J.~Keltsch, P.~Moln\'{a}r,
  Phys. Rev. E 56 (1997) 2527--2539.

\bibitem{ShChPa99} C.-R. Sheu, C.-Y. Cheng, R.-P. Pan, Phys. Rev. E 59
  (1999) 1540--1544.

\bibitem{MaSchw03} R.~Mach, F.~Schweitzer, in: W.~Banzhaf, T.~Christaller,
  P.~Dittrich, J.~T. Kim, J.~Ziegler (Eds.), Advances in Artificial Life,
  Springer, Berlin, 2003, pp. 810--820.

\bibitem{OrBaMo02b} A.~Ordemann, G.~Balazsi, F.~Moss, Physica A 325
  (2003) 260--266.

\bibitem{ErEb02} U.~Erdmann, W.~Ebeling, Fluct. Noise Lett. 3 (2003)
  L145--L154.

\bibitem{KoErSchi03} N.~Komin, U.~Erdmann, L.~Schimansky-Geier, Fluct.  Noise
  Lett. 4 (2004) L151--L159.

\bibitem{OkLe02engl} A.~Okubo, S.~A. Levin, Diffusion and Ecological Problems:
  Modern Perspectives, 2nd Edition, Springer, New York, 2001.

\bibitem{Ra94engl} J.~W.~S. Rayleigh, The Theory of Sound, 2nd Edition,
  Vol.~I, MacMillan, London, 1894.

\bibitem{St67}
R.~L. Stratonovich, Topics in the Theory of Random Noise, Vol.~II,
Gordon and
  Breach, London, 1967.

\bibitem{HaRi83} P.~H\"anggi, P.~Riseborough,  Am. J.  Phys. 51 (1983) 347--352.

\bibitem{AnAsNeVaSchi02} V.~S. Anishchenko, V.~V. Astakhov, T.~E. Vadivasova,
  A.~B. Neiman, L.~Schimansky-Geier, Nonlinear Dynamics of Chaotic and
  Stochastic Systems, Springer, Berlin Heidelberg, 2002.

\bibitem{FPE} The afore-mentioned change of variables within the set of
  Langevin equations would be equivalent to a transformation within the
  Fokker-Planck equation derived from Eq.~(\ref{langev-or}) \cite{St67,HaTh82}.

\bibitem{HaTh82} P.~H\"anggi, H.~Thomas, Phys. Rep. 88 (1982)
  207--319.

\bibitem{HTB90} P.~H\"anggi, P.~Talkner,  M.~Borkovec, Rev.
Mod. Phys. 62 (1990) 251--342.


\bibitem{HaJu95} P.~H\"anggi, P.~Jung, Adv. Chem. Phys. 89 (1995)
  239--326.

\bibitem{MiMe97} A.~S. Mikhailov, D.~Meink\"ohn, in: L.~Schimansky-Geier,
  T.~P\"oschel (Eds.), Stochastic Dynamics, Springer, Berlin, 1997, pp.
  334--345.

\bibitem{AbSt84}
M.~Abramowitz, I.~A. Stegun, Pocketbook of Mathematical Functions,
Harri
  Deutsch, Frankfurt/Main, 1984.

\end{thebibliography}

\end{document}